\documentclass[
  journal=largetwo,
  manuscript=article-type,
  year=2024,
  volume=37,
]{cup-journal}

\usepackage{amsmath}
\usepackage[nopatch]{microtype}
\usepackage{booktabs}
\usepackage{amsthm}
\usepackage{hyperref}
\usepackage{float}
\usepackage{amsmath}

\title{An Optical Daytime Astronomy Pathfinder for the Huntsman Telescope}

\author{Sarah E. Caddy}
\affiliation{Australian Astronomical Optics, Faculty of Science and Engineering, Macquarie University, Macquarie Park, NSW 2113, Australia}
\alsoaffiliation{School of Mathematical and Physical Sciences, Macquarie University, Sydney, NSW 2109, Australia}
\email[Sarah E. Caddy]{sarah.caddy@mq.edu.au}

\author{Lee R. Spitler}
\affiliation{Research Centre in Astronomy, Astrophysics \& Astrophotonics, Macquarie University, Sydney, NSW 2109, Australia}
\alsoaffiliation{Australian Astronomical Optics, Faculty of Science and Engineering, Macquarie University, Macquarie Park, NSW 2113, Australia}

\author{Simon C. Ellis}
\affiliation{Research Centre in Astronomy, Astrophysics \& Astrophotonics, Macquarie University, Sydney, NSW 2109, Australia}
\alsoaffiliation{Australian Astronomical Optics, Faculty of Science and Engineering, Macquarie University, Macquarie Park, NSW 2113, Australia}

\keywords{Optical astronomy (1776), Variable stars (1761), Artificial satellites (68), Sky brightness (1462)}

\begin{document}

\begin{abstract}
Observing stars and satellites in optical wavelengths during the day (optical daytime astronomy) has begun a resurgence of interest. The recent dramatic dimming event of Betelgeuse has spurred interest in continuous monitoring of the brightest variable stars, even when an object is only visible during the day due to their proximity to the Sun. In addition, an exponential increase in the number of satellites being launched into low Earth orbit in recent years has driven an interest in optical daytime astronomy for the detection and monitoring of satellites in space situational awareness (SSA) networks. In this paper we explore the use of the Huntsman Telescope as an optical daytime astronomy facility, by conducting an exploratory survey using a pathfinder instrument. We find that an absolute photometric accuracy between 1 - 10\% can be achieved during the day, with a detection limit of V band 4.6 mag at midday in sloan $g\textsc{\char13}$ and $r\textsc{\char13}$ wavelengths. In addition we characterise the daytime sky brightness, colour and observing conditions in order to achieve the most reliable and highest signal-to-noise observations within the limitations of the bright sky background. We undertake a 7 month survey of the brightness of Betelgeuse during the day and demonstrate that our results are in agreement with measurements from other observatories. Finally we present our preliminary results that demonstrate obtaining absolute photometric measurements of the International Space Station during the day. 
\end{abstract}

\section{INTRODUCTION}\label{the_start}
For centuries, Astronomers have been captivated by the idea of observing stars during the day from the ground, in visible wavelengths. Attempts to do so using nothing but the human eye date back as far as the fourth century B.C, where Aristotle postulated that ``people in pits and wells sometimes see the stars" \citep{aristotle_generation_1942}. Ever since, perhaps spurred on by the numerous recounts of Aristotle's musings throughout history \citep{richard_observing_1992}, Astronomers have been reportedly found disappointed at the bottom of chimneys and caverns hoping to capture a glimpse of a star in broad daylight \citep{hughes_seeing_1983}. These were the first attempts at optical astronomy during the day.
\\
\\
Interest in the accurate photometry of objects observed during the day is driven by two applications. Space Situational Awareness (SSA) for the purposes of identification and monitoring of satellites, and the continuous monitoring of bright variable ecliptic stars, such as Betelgeuse and Aldebaran.

\subsection{Space Situational Awareness}
\begin{figure*}[ht!!!]
    \centering
    \includegraphics[width=15cm]{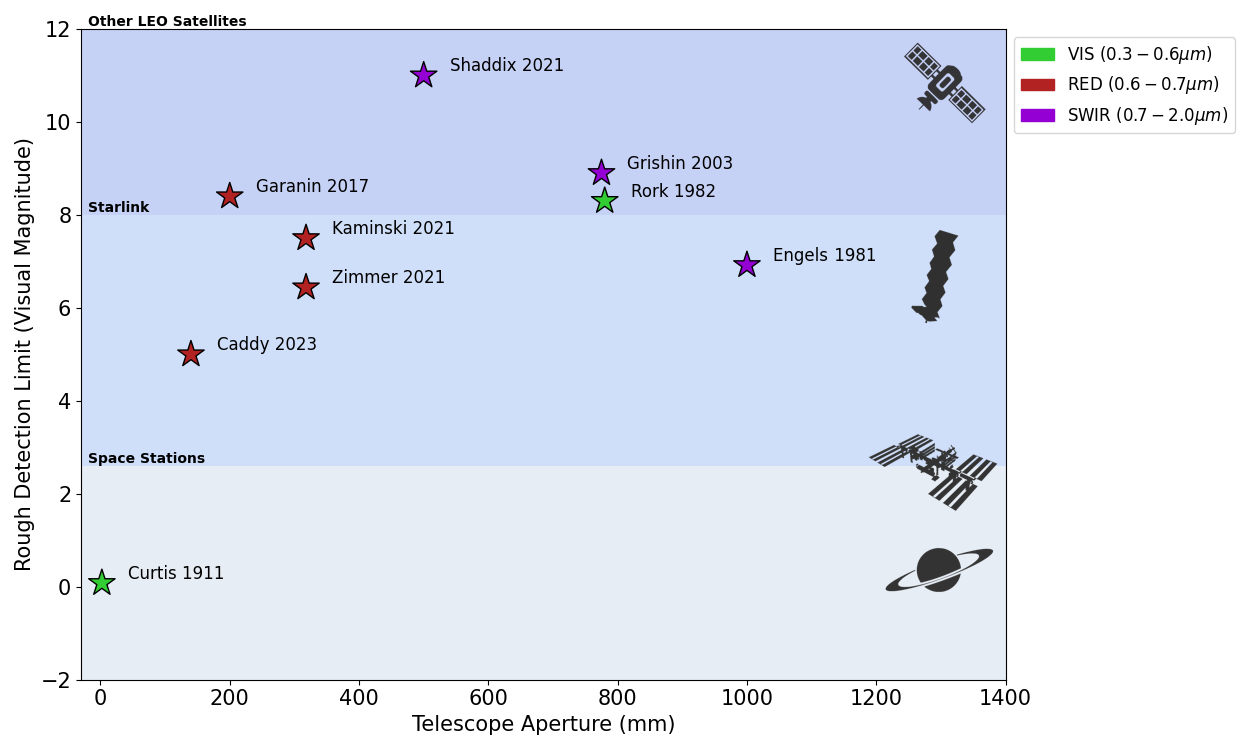}
    \caption{A sample of detection limits for visual (green), red (red), and short wave infrared (SWIR) (purple) presented in the literature for daytime observations of stars. The data included is from the works of \cite{curtis_seeing_1911, engels_infrared_1981, rork_ground-based_1982, grishin_infrared_2003, garanin_daytime_2017, kaminski_optimizing_2021,shaddix_daytime_2021, zimmer_overcoming_2021} and this work. Only works that explicitly mention a detection limit are included. Due to the lack of public information surrounding some of the hardware used in daytime observing by various research groups, a reliably reported parameter of telescope aperture is plotted against the detection limit. It should be noted that each of these instruments likely have very different focal lengths and other characteristics, as well as different data reduction methods that may bias results. Three shaded regions in visual magnitudes are also presented, showing rough predicted and observed magnitude lower margins during the day. The brightest stars and planets are visible to most optical systems including the eye \citep{curtis_seeing_1911} during the day, and are around $\sim0th$ magnitude. The International Space Station has been observed to have a visual magnitude during the day of $\sim 2nd$ magnitude at zenith (this work). Starlink satellites have been observed in the day up to 3.0 mag  \citep{halferty_photometric_2022} at night, and up to 2.6 mag during the day \citep{zimmer_overcoming_2021}. Following the work of \cite{kaminski_optimizing_2021}, most other LEO satellites have a visual magnitude fainter than 8th magnitude.}
    \label{fig:state_of_play}
\end{figure*}

Research into viewing artificial satellites during the day has been explored in the last few decades, motivated by an increasing need for SSA \citep{shaddix_daytime_2021, estell_daylight_2019, skuljan_photometric_2018, thomas_daytime_2017, roggemann_daytime_2010, rork_ground-based_1982}. Distributed facilities like the Falcon Telescope Network \citep{chun_falcon_2014} are playing an increasingly important role monitoring the sky now that popular orbits such as Geostationary, and Low Earth Orbits (LEO) have become over crowded \citep{barentine_aggregate_2023}. In the next 10 years alone, over 50,000 LEO satellites are planed for launch \citep{zimmer_overcoming_2021} in addition to the estimated $>1,000 ,000$ 1~cm sized debris that are of risk to LEO satellites \citep{barentine_aggregate_2023}. The need for facilities that can constantly monitor the sky came into the public view in 2009, when two satellites - one an Iridium communications satellite and one a decommissioned Russian military satellite - collided and produced a dangerous cloud of debris travelling and thousands of meters per second in LEO \citep{gasparini_space_2010}. The event could have been avoided, if the Iridium satellite had warning from SSA facilities to manoeuvre out of the collision course.
\\
\\
Traditional observations of objects in LEO are limited to terminator illuminated conditions. This occurs during in early evening and morning when the observing location is dark, and the object in LEO is directly illuminated by the Sun \citep{zimmer_optimizing_2020}. In addition, the time at which a satellite may be overhead at a typical observing location during terminator illuminated conditions may only be a few minutes - if at all - and may only be observable every few weeks from a given location on the Earth \citep{zimmer_overcoming_2021}. Earthshine is defined as light scattered and remitted from the surface of the Earth, and has been shown to be an important contribution to satellite optical brightness \citep{fankhauser_satellite_2023}. At $\sim 0.3 \text{-} 2 \mu m$, Earthshine is dominated by reflected sunlight \citep{caddy_toward_2022} and predominantly bluer in colour. In observing conditions where the optical upwards flux from Earthshine is the strongest, typically in conditions where there are large cloud systems or regions of ice below a satellite in LEO, satellites can be illuminated comparably to terminator illuminated conditions \citep{zimmer_optimizing_2020}. By taking advantage of daytime passes that are illuminated in the direction of the observer by Earthshine \citep{zimmer_optimizing_2020}, the amount of time per day a typical satellite can theoretically be observed is increased from $\sim 11\%$ to $\sim 56\%$ \citep{estell_daylight_2019}. The use of an optimised ``cathemeral" telescope network (active in both day and night conditions) such as that described in \cite{shaddix_daytime_2021, shaddix_daytime_2019}, may see mean improvements on observable satellite passes of $300-400\%$, to as much as $1000\%$ for some objects as opposed to operating only during terminator illuminated conditions. 
\\
\\
In addition to \cite{zimmer_overcoming_2021, shaddix_daytime_2021, kaminski_optimizing_2021, zimmer_optimizing_2020, estell_daylight_2019} who demonstrate isolated detections in optical and shortwave infrared daytime satellite observations, there have been some attempts at simply detecting an object during the day for the purposes of exploring daytime SSA, or even just curiosity. An early attempt by \cite{curtis_seeing_1911} reports to have observed the brightest stars by eye, by calculating their expected position with relation to architectural landmarks. \cite{grishin_infrared_2003} reports observations of 8.9 mag stars using an IR camera during the day. \cite{garanin_daytime_2017} reports detecting V band 7th and 8th magnitude stars using a 7.9 inch f/10 telescope and a video camera for the purposes of SSA but made no attempt at photometry. Using a 12.5 inch f/7 telescope for SSA \cite{zimmer_optimizing_2020, zimmer_overcoming_2021} report the detection of 7th magnitude stars. In addition, there has also been work towards improving daytime sky brightness models in order to create accurate exposure time calculators for SSA \citep{thomas_daytime_2017, roggemann_daytime_2010}. Some work has also been done to explore the possibility of observing geostationary satellites during more favorable times of the day where the sky brightness is fainter. \cite{cognion_large_2013} undertook a survey of geostationary satellites at large phase angles at night in order to produce empirical models of target brightness for the purpose of observing these targets in early morning and late afternoon. \autoref{fig:state_of_play} summarises the rough detection limit that observations in the literature have achieved. These estimates are plotted as a function of the telescope aperture, which is the most reliably referenced characteristic of the observational set up. The detection limit is an important parameter that impacts the productivity of a cathemeral telescope as it sets the limit on the number of SSA targets that can be monitored during the day \citep{zimmer_overcoming_2021}. \cite{shaddix_daytime_2021} reports the faintest daytime detection limits of roughly 11th magnitude using the Aquila telescope at SWIR wavelengths. 
\\
\\
In addition to detecting and tracking satellites, photometric observations are a useful tool that is currently being explored and developed for SSA applications \citep{pearce_examining_2019,skuljan_photometric_2018, schmitt_multicolor_2016,frith_nir_2015, cognion_large_2013, scott_small-aperture_2009, moore_photometric_1959}. Accurate multi-wavelength colour observations can provide constraints of the composition of a satellite or debris and aid in its identification, as well as the composition of different components of a single satellite. In the work of \cite{schmitt_multicolor_2016} colour differences are found to be as great as $V-I \sim 1.2$ mag and $B-V \sim 0.75$ mag and are split into two distinct groupings separated in a $V-I$ verses $B-V$ diagram by 0.2 mag. This distinct difference in colour could be explained by the use of gold verses kapton, and indicates the satellites made of different material have different spectral signatures that can be used to identify them. The brightness of a satellite as a function of phase angle can also be used to identify one satellite bus from another in cluster type configurations in GEO. The work of \cite{scott_small-aperture_2009} has led to a method of determining miss-classification of objects in NORAD catalogs. In addition, some evidence has arisen of the reddening of satellites with age due to space weathering \citep{pearce_examining_2019}. Monitoring of this reddening effect may be useful as a tool to assess the age and condition of a satellite in orbit.  
\\
\\
There is a clear need for dedicated cathemeral telescope networks that have the capability of detecting and tracking satellites during the day and night. Despite this need, only a handful of facilities around the world have begun testing operations as SSA facilities during the day and no facilities that the authors are aware of operating in Australia (although some have been proposed \citep{shaddix_daytime_2021}) - which is a key strategic location for SSA \citep{vignelles_australian_2021}. Of the facilities that have demonstrated isolated cases of successful daylight satellite detection in the optical and shortwave infrared, none of these facilities operate autonomously for the purposes of detecting, tracking and identifying satellites during the day. In addition, there have been no accurate photometric observations of satellites conducted during the day. This presents an ideal opportunity to fulfill this need and utilise the Huntsman Telescope as a cathemeral telescope facility.

\subsection{Variable Star Monitoring}
In addition to these practical applications, optical daytime astronomy is also of increasing interest due to recent activity of the red supergiant star Betelgeuse. Regular photometric observations of Betelgeuse in the optical are recorded by organisations such as the American Association of Variable Star Observers (AAVSO). However for a period of about 4 months every year the star is located too close to the Sun and not observable at night, resulting in a significant annual gap in the light curve \citep{nickel_daylight_2021}. The star is considered a semi-regular variable, but from November 2019 to April 2020 Betelgeuse was observed to experience a large and rapid dim in brightness from V band $\sim0.5$ to a historic minimum of $1.614 \pm 0.008$ \citep{dupree_great_2022, kravchenko_atmosphere_2021, montarges_dusty_2021, montarges_great_2020, guinan_fainting_2019}. The dimming event has been attributed to surface mass ejection as well as changes in the temperature of the photosphere \citep{dupree_great_2022, kravchenko_atmosphere_2021, montarges_dusty_2021}. Continued regular observations have since confirmed the disappearance of the $\sim400$ day pulsation period in the optical and radial velocity \citep{dupree_great_2022}. Betelgeuse presents an excellent opportunity to observe a star up close in its final phases of life. The physics behind Betelgeuse's photometric variability and mass loss is still poorly understood \citep{kravchenko_atmosphere_2021, montarges_dusty_2021}, and optical daytime facilities offer an opportunity to continue to monitor stars like Betelgeuse uninterrupted, year round.  
\\
\\
There have been several attempts to perform accurate optical daytime photometry for the purpose of bright star monitoring. \cite{engels_infrared_1981} reports a photometric error of 0.03 mag using a photometer on the 1-meter telescope and ESO - La Silla. \cite{miles_daytime_2007} reports a photometric error of V band $\pm 0.03$ mag observing Betelgeuse using a 60mm refactor, and a Starlight-Xpress CCD. To avoid saturation due to the longer exposure time limit of the CCD, a 1\% neutral density filter is used in combination with a green V band filter. Following the dimming event of Betelgeuse, \cite{nickel_daylight_2021} performed optical daytime photometry from the ground to capture a continuous light curve during the months the star was only accessible during the day. A 7.9 inch f5 Newtonian telescope was used in a backyard in Mainz, Germany with an astrophotography CCD camera. A $1\%$ neutral density filter was also added in later iterations. Extinction parameters were calculated daily using bright reference stars, and found that extinction strength was correlated with haze content in the sky. The set up achieved a photometric accuracy for Betelgeuse of V band $0.02\pm0.008$~mag to $0.04\pm0.013$~mag depending on the sun separation distance \citep{nickel_daylight_2021}. Their absolute magnitude measurements are in agreement with observations made from space using the STEREO Solar telescope \citep{dupree_photometry_2020}. These results demonstrated the possibility of achieving $\sim 1\%$ photometry during the day to monitor variable bright stars that are inaccessible to traditional ground based optical observatories for periods of time each year. 

\subsection{The Huntsman Telescope}
\begin{figure*}[ht!!!]
    \centering
    \includegraphics[width=\linewidth]{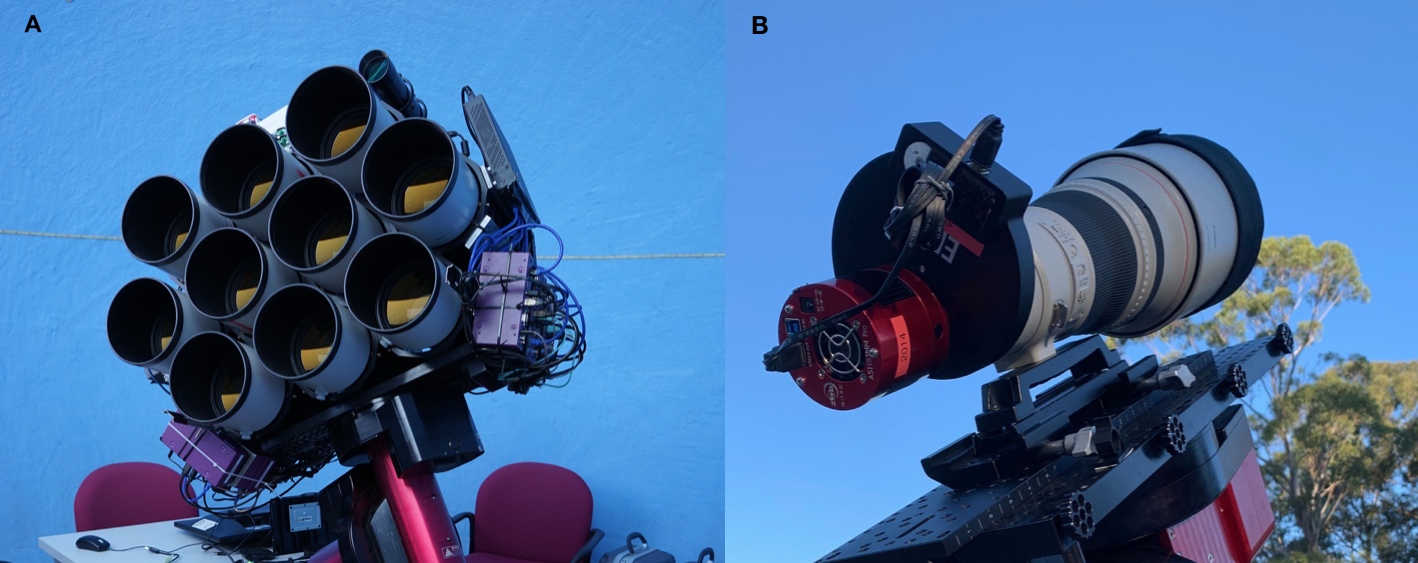}
    \caption{A) The Huntsman Telescope remote observing facility located at Siding Spring Observatory, Australia. The telescope consists of 10 Canon 400~mm f/2.8 lenses in an array configured to cover the same field of view of $1.89^\circ \times 1.26^\circ$ with a pixel scale of 1.24''. B) The pathfinder instrument used to test daytime observing modes for this work consisting of a single lens. The unit is located at Macquarie University Observatory, Sydney, Australia} 
    \label{fig:huntsman}
\end{figure*}

In this work, we explore the capabilities of The Huntsman Telescope to observe objects during the day. The Huntsman Telescope (\citealp{spitler_huntsman_2019}; see \autoref{fig:huntsman}A) is a $<$0.5~m class research facility located at Siding Spring Observatory, Australia on Gamilarray, Wiradjuri and Wayilwan country. The telescope consists of 10 Canon 400~mm f/2.8 lenses in an array configured to cover the same field of view of $1.89^\circ \times 1.26^\circ$ with a pixel scale of 1.24\textquotesingle\textquotesingle. The telescope builds on the design of the Dragonfly Telephoto Array \citep{abraham_ultralow_2014} as inspiration, with the primary research goals being low surface brightness imaging and supporting optical transient discovery for the Deeper Wider Faster program \citep{andreoni_deeper_2017}.
\\
\\
The Huntsman telescope facility addresses many of the challenges experienced with optical daytime astronomy. Ultimately, the sensitivity and productivity of an optical daytime facility is dependant on the field of view, and the sky background intensity per pixel. Fast telescopes with high \'etendu are optimal for this observing mode. At f2.8, Huntsman's large field of view is ideal for SSA observations to ensure a target is within an image given an uncertain position. Following an upgrade in 2020, the telescope is equipped with ZWO-brand, ASI183MM Pro cameras that use CMOS sensor technology. These sensors are capable of down to 32 microsecond duration exposures and high frame rates of up to 271 fps in windowed read out modes. At this high frame rate for bright star monitoring, we expect the instrument to be capable of sampling the speckled Airy disk due to scintillation, with each speckle remaining diffraction limited. The capability of the instrument to take photometric data in up to 5 different bandpasses simultaneously in its current configuration is useful for the potential to monitor colour information of tumbling, moving SSA targets where simultaneous multi-wavelength data are required. The multi-lens design may also aid in reducing false positive detections. Each lens, focuser, filterwheel and camera unit is controlled by a dedicated edge computing unit, the Jetson Xavier, with capability of onboard real time processing. This will enable the telescope to manage large data streams while operating at high frame rates - a known problem for SSA facilities operating under similar conditions \citep{zimmer_overcoming_2021, shaddix_daytime_2019}. The telescope is able to be left for weeks at a time with pre-programmed scheduling, equipped with a dedicated weather station and robotic dome control.
\\
\\
In this work we describe a Huntsman Telescope pathfinder instrument located at Macquarie University observatory (see \autoref{fig:huntsman}B) that is used to explore the capability to produce photometric observations taken during the day using the Huntsman Telescope. We present our methods of data reduction, some of the challenges of observing during the day, and identify the limitations of the system under different environmental conditions. We present the total photometric accuracy of the system operating during the day following an initial 7 month survey, and describe our planned future work and ongoing daytime observations of bright ecliptic variable stars. We explore the photometric accuracy of observing during the day utilising bright stars, for the purposes of applying these techniques to bright star monitoring. In addition we demonstrate this use case using the variable star Betelgeuse. Finally we share preliminary results of photometry of the International Space Station (ISS) during the day. While photometry of a large sample of satellites is out of the scope of this work, we utilise the results of these first tests to explore the feasibility of upgrading Huntsman to perform such observations.
 
\section{Observations}\label{sec:method}
Optical astronomy performed during the day is a largely undocumented field of observational astronomy. As a result, the observing techniques needed to acquire reliable photometric data were developed over a year long period through a process of trial and error. Initial observation were conducted with the Huntsman Telescope in April of 2021, but were not rigorously tested until the later half of 2022 when deployed on the Huntsman Telescope pathfinder. Finally in February of 2023 a formal survey of bright stars was undertaken to determine the photometric accuracy of an optical daytime facility. In addition, observations of fainter stars down to V band $\sim 6$ mag were also conducted to test observation limits. Here we describe the survey and tests undertaken to characterise the performance of the hardware, refine the observing techniques, develop the data reduction pipeline and determine the absolute photometric accuracy of the system during the day. 
 
\subsection{Hardware and Setup}\label{obs_set}
Daytime astronomy is a new, experimental observing mode for the Huntsman Telescope. The telescope's remote location poses complications for the rapid deployment of experimental observing techniques and so order to efficiently iterate on survey design and reduce any risk in damaging the Huntsman Telescope facility, this work is performed at a local controlled environment. We use a pathfinder instrument in the form of a single, identical Huntsman lens unit at the Macquarie University Observatory in Sydney, Australia. 
\\
\\
The Huntsman Telescope pathfinder, (mini-Huntsman herein) consists of one Canon 400mm f2.8L lens, a ZWO ASI18300MM Pro camera, and an astromechanics focuser (see \autoref{fig:huntsman}B). Mini-Huntsman is mounted on a Software Bisque ME2 mount with a ZWO filter wheel with Sloan $g$\text{\textquotesingle}, and $r$\text{\textquotesingle} broadband filters, as well as SII and H$\alpha$ narrow band filters. Observations are conducted with the narrow band filters, but are not used in photometric observations and will be the subject of future investigations. The telescope is covered in a weatherproof thermal blanket for storage, and is not located in a dome. For these tests, no sun shields are used apart from the baffle that is supplied with the lens. No guide scope and camera are used in these tests. The telescope is operated manually via TheSkyX desktop interface, and ZWO software ASICap is used to take data. Throughout these tests the camera is cooled to $0^\circ C \pm 0.5 ^\circ C$. 

\subsection{Description of the Survey} \label{surv_1} 
\begin{figure}[ht!!!]
    \centering
    \includegraphics[width=\linewidth]{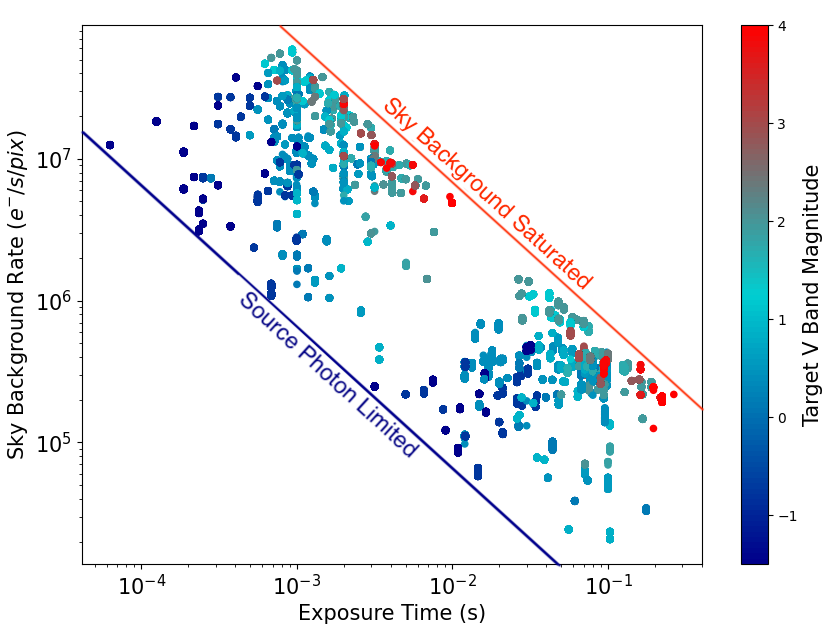}
    \caption{The observed sky background rate plotted as a function of the exposure time for all filters and targets. The colour bar illustrates the catalog V band magnitude of the target observed. On the far right, for higher exposure times the image becomes saturated due to the high sky background. To the far left observations are limited by the photon count from the source. Two clear clusters are formed. The lower right are the narrowband filters with much higher exposure times due to their narrow bandwidth, and the upper left are the broad band observations. Exposure times are tuned per every set of observations to ensure that either the image if not over saturated by the target, or by the sky background.}
    \label{fig:exp_times}
\end{figure}

The first observations were conducted throughout March 2023 at the Macquarie University Observatory site. Observations are made for 35 different stars of varying colour and magnitude, and spanning different times of the day and varying environmental conditions. Stars that are bright for reliable high SNR measurements that are used as reference stars for differential photometry during the day are also often variable stars themselves, and so we limit photometric reference stars to targets that do not vary significantly. Daytime temperature estimates and sky conditions are recorded daily using Macquarie Observatory's weather mast. For the first part of the day, the telescope is subjected to direct sunlight. From mid afternoon onward (depending on the time of year) the telescope is shaded by nearby trees. This exposure to the elements causes the instrument to undergo large temperature variations. Due to the large temperature fluctuations during the day, the instrument is refocused for every target. As more temperature data is collected, this may form the basis for a focus offset algorithm, if offsets are found to be consistent with temperature in a future work. 
\\
\\
Exposure times are set manually to ensure there is no saturation of the target, or of the sky background. \autoref{fig:exp_times} illustrates the exposure time used for each target as a function of the detected sky background rate, for all filters used in the work. The colour bar shows the catalog V band magnitude of the target. Observations to the right of the plot are more likely to be limited in exposure time by the saturation of the sky background depending on the sky brightness and filter used. Observations to the left of the plot are more likely to be limited by the saturation of the target. The narrowband filters occupy the cluster to lower right hand side of the plot as their narrow bandwidth as opposed to the broad band filters allow for a higher exposure time, while broadband observations occupy the upper right cluster. As a general rule, the exposure time is tuned on a per exposure set bases to ensure that the exposure time is as high as possible without saturation.
\\
\\
Scintillation noise due to the Earths turbulent atmosphere at high altitudes, is a dominant source of noise during the day. For short exposures where the exposure time approximates the atmospheric coherence time, the scintillation regime that is probed is 'short' in which the intensity speckles from the target star appear frozen, and no temporal averaging occurs \citep{osborn_atmospheric_2015}. For mini-Huntsman with a small aperture of 140mm, the change between long and short scintillation regimes is a few hundredths to a few tenths of a second \citep{osborn_atmospheric_2015}. As illustrated in \autoref{fig:exp_times}, broad band filters probe this short regime from exposure times of $10^{-4} - 10^{-2}$ seconds. Narrow band exposures however, at exposure times of $10^{-2} - 10^0$ seconds, intensity speckles begin to temporally average. To take advantage of the short scintillation regime in broadband filters, a relatively small Full Width Half Maximum (FWHM) for the {\tt photutils} Gaussian kernel of 3 pixels is used to convolve the images before source detection, favouring small scale structure in short broadband exposures. We do not consider the process of lucky imaging in this work, where by only serendipitous moments of high image quality are selected to be analysed or stacked. This allows us to sample a broad parameter space of observing conditions in order to determine the best practices in this exploratory work, however, will be considered in future refined work.
\\
\\
The telescope is focused both using multiple iterations of TheSkyX auto focus procedure at the start of the observing session for each filter, and then manually adjusted for every target as the ambient temperature varies throughout the day, and filter offsets are calculated. Care is taken to ensure the target is close to the centre of the field of view. The CMOS detector is read out using a central sub-array of $320 \times 240$ pixels centered on the target to ensure the highest frame rates can be achieved, typically $>100$~fps for broad band exposures, and is limited by the data rates of USB3. The gain is set to 0 to maximise the pixel well depth for every exposure such that for the ZWO ASI183MM Pro, $1\,ADU \simeq 3.88 e^{-}$. We do not change the gain value for the duration of this entire work, and use of higher gain values may be explored in a future work. For every target and every filter, 1000 consecutive images are taken. Data were taken during brief periods of high cloud or haze as well as particularly low target altitudes, and observations during high winds. These data are labeled and used to aid in determining data quality metrics for the data reduction pipeline in future works. In addition, fainter stars of a variety colours and magnitudes were also targeted to determine the limiting magnitude of the system under different environmental conditions. Following the first exploratory observations, the telescope was mounted permanently to the pier, and a refined survey was conducted from April - August 2023 using the results of the first. A new pointing model was created to ensure the target is centered for every exposure. We attempt to take data only on days when the sky is absolutely clear, and there is little wind to ensure mechanically stable observing conditions, however this was not possible for some dates during the Winter period of June - August which were plagued by poor weather conditions. Only bright stars $<$ 3rd magnitude V band are used in order to derive the photometric zeropoint, extinction coefficient, colour terms, as well as statistical and photometric uncertainties for the system.
\\
\\
In addition to the primary goals of this work, we also carry out continuous observations of Betelgeuse, along with several reference stars. These observations are conducted approximately 2 -- 3 times per month throughout the year. Observations are taken in $r$\text{\textquotesingle}, and $g$\text{\textquotesingle}, with an aim of 3+ pointings per day. This ensures that the extinction coefficient can be monitored from day to day, which was demonstrated to be a critical aspect of optical daytime photometry in the work of \cite{nickel_daylight_2021}. The collected data is presented and compared with AAVSO observations to demonstrate the system's potential as an optical daytime facility. 
\\
\\
Finally, we explore the initial capabilities of the mount to target and track artificial satellites for SSA. The satellite tracking feature of TheSKyX is used for these tests, with ephemeris sourced from Celestrak. The $g$\text{\textquotesingle} filter is used for these initial tests to maximise the SNR of the satellite, following the results of \cite{zimmer_optimizing_2020, zimmer_overcoming_2021}. For these initial tests, the International Space Station (ISS) is tracked and imaged during the day, and a $g$\text{\textquotesingle} calibrated light curve is computed for the pass. Detection and tracking of other satellite targets is out of the scope of this work, and is the subject of ongoing tests. 

\subsection{Data Reduction Pipeline}
The data reduction pipeline for this work makes use of {\tt photutils} for source detection and aperture photometry. Gaussian fitting is performed using {\tt Source Extractor}. A new algorithm was created that checks for the presence of a single source in the data iteratively, starting from $6 \times$ the standard deviation of the sky background to 1 in units of 0.5. Should the algorithm detect multiple sources for a single threshold value, the exposure is rejected. This is required because of the large number of false positive detections originating from terrestrial objects, which have been dubbed `angels' for their intermittent appearance \citep{rork_ground-based_1982}. These objects are reported to vary in height from 30m to 2km and may include seed packets, insects, or ice crystals. The probability of an angel being present in an image increases as the sun separation distance is reduced. After the set of 1000 consecutive images is processed for single sources, the algorithm determines the detected source centroid and the median location of the target across the 1000 images. If any single source is located more than 5 standard deviations in $x$ or $y$ from this median, it is rejected. The FWHM is calculated using two methods. This is done due to the large range of distorted PSF shapes, particularly during periods of high scintillation or poor seeing, and for fainter targets. Each are compared to determine their accuracy and reliability. One method estimates the FWHM using the {\tt Source Extractor} Gaussian fit a and b parameters: 
\begin{equation*}
    FWHM = 2 \sqrt{(\ln(2) \times (a^2 + b^2))}
\end{equation*}
In addition, we calculate the FWHM by flattening a $20 \times 20$ postage stamp of the target in both axes and fitting a 2D Gaussian using {\tt Scipy} to ensure the PSF wings are properly sampled. The final FWHM is then the average of the two axes. We find that this method is more reliable and accurate than {\tt Source Extractor}, which tends to over estimate the PSF FWHM for daytime sources, and so is not used in this work.
\\
\\
The increased atmospheric turbulence during the day causes the PSF FWHM to vary significantly. For this reason an aperture size for photometry was selected so that large enough that the probability of losing flux is low. While this will decrease the SNR, it also removes the need for individual aperture corrections for every exposure. For all photometry, we adopt an aperture radius that meets this criteria: the radius in which the derivative of the target aperture flux (background subtracted) as a function of aperture radius approaches the increase in noise for the $n+1$ aperture radius where $n$ is in units of pixels. We find that the aperture radius that satisfies this criteria is $\sim 11$ pixels. This aperture size is used for the reduction of all data in this work. This ensures we minimise the flux lost in the wings of observations that are heavily impacted by scintillation, and removes the need for aperture corrections. In future works, a SNR-optimised algorithm will be developed to determine the best aperture size as a function of the seeing conditions and perform aperture corrections where needed. Each set of 1000 consecutive exposures is given a unique hexadecimal ID, and each exposure is given a unique number in addition to the hexadecimal set ID so they can be easily tracked. The sky rate is calculated as the median sky background in an annulus around the source in $ADU/pix/s$. The total flux from the source is then calculated as the sum of the background subtracted flux in an aperture in $e^{-}/s$. The final instrumental magnitude is then:
\begin{equation*}
m_{inst}=-2.5\log_{10}\Big(\frac{\sum{f_{r}}/t} {\int_{\lambda_{1}}^{\lambda_{2}}{TP \times QE}}\Big)
\end{equation*}
Where $\lambda_{2} - \lambda_{1}$ is the bandwidth of the filter, $TP$ is the throughput of the filter, $QE$ is the quantum efficiency of the detector, $t$ is the exposure time in seconds, and $f_{r}$ is the sum of background subtracted flux in $e^{-}$ in an aperture of radius $R$. Catalog $r$\text{\textquotesingle} and $g$\text{\textquotesingle} magnitudes used to calibrate mini-Huntsman sources saturate at $\sim 14$ mag \citep{york_sloan_2000} in the Sloan Digital Sky Survey and $\sim 13$ mag in the SkyMapper Southern Sky Survey, so in order to compare mini-Huntsman magnitudes of bright stars accessible during the day we use transformation equations from Johnson-Cousins U, B and V to Sloan $g$\text{\textquotesingle} and $r$\text{\textquotesingle} presented in \cite{jester_sloan_2005}. For stars that satisfy $U-B < 0$:
\begin{equation*}
    g\text{\textquotesingle} = V + 0.64\times(B-V) - 0.13 \pm 0.01
\end{equation*} 
\begin{equation*}
    r\text{\textquotesingle} = V - 0.46\times(B-V) + 0.11 \pm 0.03  
\end{equation*}
And for stars where $U-B \geq 0$:
\begin{equation*}
    g\text{\textquotesingle} = V + 0.60\times(B-V) - 0.12 \pm 0.03
\end{equation*} 
\begin{equation*}
    r\text{\textquotesingle} = V - 0.42\times(B-V) + 0.11 \pm 0.03  
\end{equation*}
Johnson-Cousins U, B and V are retrieved from the Simbad database using the \texttt{astropy} astroquery method. 
\\
\\
Mini-Huntsman photometric data has been calibrated to the Sloan photometric system using the following system of photometric zeropoint equations:
\begin{equation*}
         0 = r\text{\textquotesingle}_{inst} - r\text{\textquotesingle} - ZP_{r\text{\textquotesingle}} - C_{r\text{\textquotesingle}}(g\text{\textquotesingle} - r\text{\textquotesingle}) - k_{r\text{\textquotesingle}}(X)   
\end{equation*}
\begin{equation*}
     0 = g\text{\textquotesingle}_{inst} - g\text{\textquotesingle} - ZP_{g\text{\textquotesingle}} - C_{g\text{\textquotesingle}}(g\text{\textquotesingle} - r\text{\textquotesingle}) - k_{g\text{\textquotesingle}}(X)     
\end{equation*} 
Where $r\text{\textquotesingle}_{inst}$ and $g\text{\textquotesingle}_{inst}$ are the instrument magnitudes in units of $e^{-}/s$, for an aperture of radius $R$, after correcting for the measured throughput and QE. $r$\text{\textquotesingle} and $g$\text{\textquotesingle} are the catalog magnitude for the target star, $ZP_{r\text{\textquotesingle}}$ and $ZP_{g\text{\textquotesingle}}$ are the system zeropoints, $C_{r\text{\textquotesingle}}$ and $C_{g\text{\textquotesingle}}$ are colour indices, $k_{r\text{\textquotesingle}}$ and $k_{g\text{\textquotesingle}}$ are the atmospheric extinction coefficients, and $X$ is the airmass at the time of the measurement. The two equations can be solved through a least squares approach to determine the coefficients. The colour indices are determined using data for a single day with the largest sample of stars with varying colours and magnitudes. The extinction parameter and zeropoint is calculated daily to accommodate for varying atmospheric conditions. We test the photometric calibration using two methods. Firstly, a least squares approach is implemented using \texttt{Scipy}. In addition, to identify any possible degeneracy within the fit a Markov Chain Monte Carlo (MCMC) approach is also implemented using \texttt{PyMC3}. In this method, the model is trained 10 times, where the uniform priors for each parameter are updated with the median of the $n-1$ fit. The initial guess for each prior is $0$. For all models, the \texttt{PyMC3} and \texttt{Scipy} methods produce the same final results for parameter estimation, with the difference being the sacrifice of speed (\texttt{Scipy}), for parameter space visualisation (\texttt{PyMC3}). Following these results we use the \texttt{Scipy} method throughout this work in favour of computation time. 

\section{Results}

As we explore the capabilities of the Huntsman Telescope to provide accurate photometry of bright targets during the day, we aim to answer three questions. 1) What is the absolute, reliable photometric accuracy achievable during the day? 2) What factors influence daytime photometric accuracy? And 3) what impacts the Huntsman Telescope's productivity as an observatory operating during the day. Finally we present our light curve of the test target Betelgeuse over a period of 7 months, and preliminary work to detect satellite passes during the day. 

\begin{figure*}[!hbt]
    \centering
    \includegraphics[width=\linewidth]{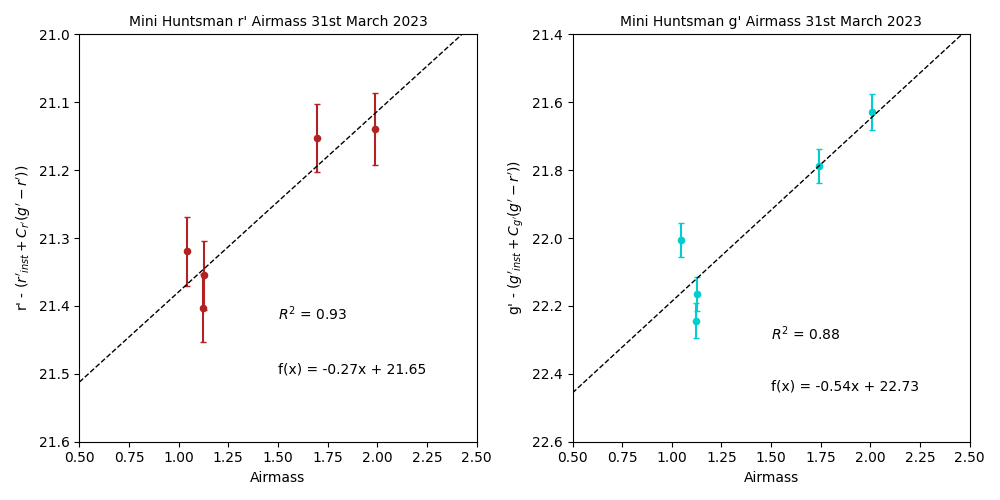}
    \caption{An example airmass plot for the 31st of March 2023, showing the calculated magnitude as a function of airmass for 5 bright reference stars. Observing logs show a clear day, no wind, with tops of $25^{\circ}C$ ($77^{\circ}F$). Each data point is the median of a set of 1000 exposures for that reference star. The extinction coefficient for this day is found to be 0.27 for $r\text{\text{\textquotesingle}}$ and 0.54 for $g\text{\text{\textquotesingle}}$ with zeropoints of 21.65 and 22.73 respectively. The error bars encapsulate the statistical error of the set of 1000 observations, and the flat field error in quadrature. The Chi squared values for the linear fit are calculated to illustrate the fit, with a higher Chi squared of 0.93 for $r\text{\text{\textquotesingle}}$ as opposed to $g\text{\text{\textquotesingle}}$. Scatter about an airmass of 1 may be due to dust in the air produced by lawn mowing that observing logs showed to occur close to the observatory at the time of observations.}
    \label{fig:airmass_plots}
\end{figure*}

\label{phot_acc}
\subsection{Photometric Accuracy of Observations}
We present the photometric accuracy achieved for our preliminary observations of various standard stars over the course of our exploratory work with the Huntsman Telescope Pathfinder. The total photometric error is calculated for days where the observing logs do not mention significant impact due to cloud cover and bushfire smoke, which impacted 3 of the days in which observations took place.
\\
\\
A significant source of error in our preliminary results was due an inability to calibrate our images with flat fielding data. Due to a software fault, the centroid of the windowed sub-array ($320 \times 240$ pixels of $5496 \times 3672$) was not captured in observations. In addition, a relatively poor pointing model resulted in stars moving in the FOV from pointing to pointing. Due to these issues, the images could not be corrected pixel-to-pixel sensitivity variations with flat field data as a result of the software fault. The Canon lenses introduce a radially-symmetric vignetting pattern such that the outer regions of the full image frame have 27\% less light compared to the centre due to optical vignetting. To quantitatively estimate the impact of this issue on typical data, a mean radial profile of a full-frame normalised $r\text{\text{\textquotesingle}}$ band flat is produced and applied to a typical Betelgeuse image. We find that the flat field error due to the mini-Huntsman vignetting pattern introduces a maximum possible error of $\sim 0.2 mag$ for targets on the extreme outer edge of the image, compared to the central pixel. For targets within $<1000$ pixels of the central pixel (the most plausible error) this is reduced to $\sim 0.05$ mag. It should also be noted that the number of observed stars vary significantly from day to day. The factors that influenced the number of stars observed were available time and the weather. Patchy cloud and frequent late afternoon dense cloud cover impacted the number of stars observed. Days vary between 20+ targets utilised for determining performance limits, to 3-5 specifically for differential photometry. On some days, only the star of interest, Betelgeuse, could be observed with no calibration stars. For these days, the average sky extinction coefficient for the entire observing run is used to calibrate the data.
\\
\begin{figure*}[ht!!!]
    \centering
    \includegraphics[width=\linewidth]{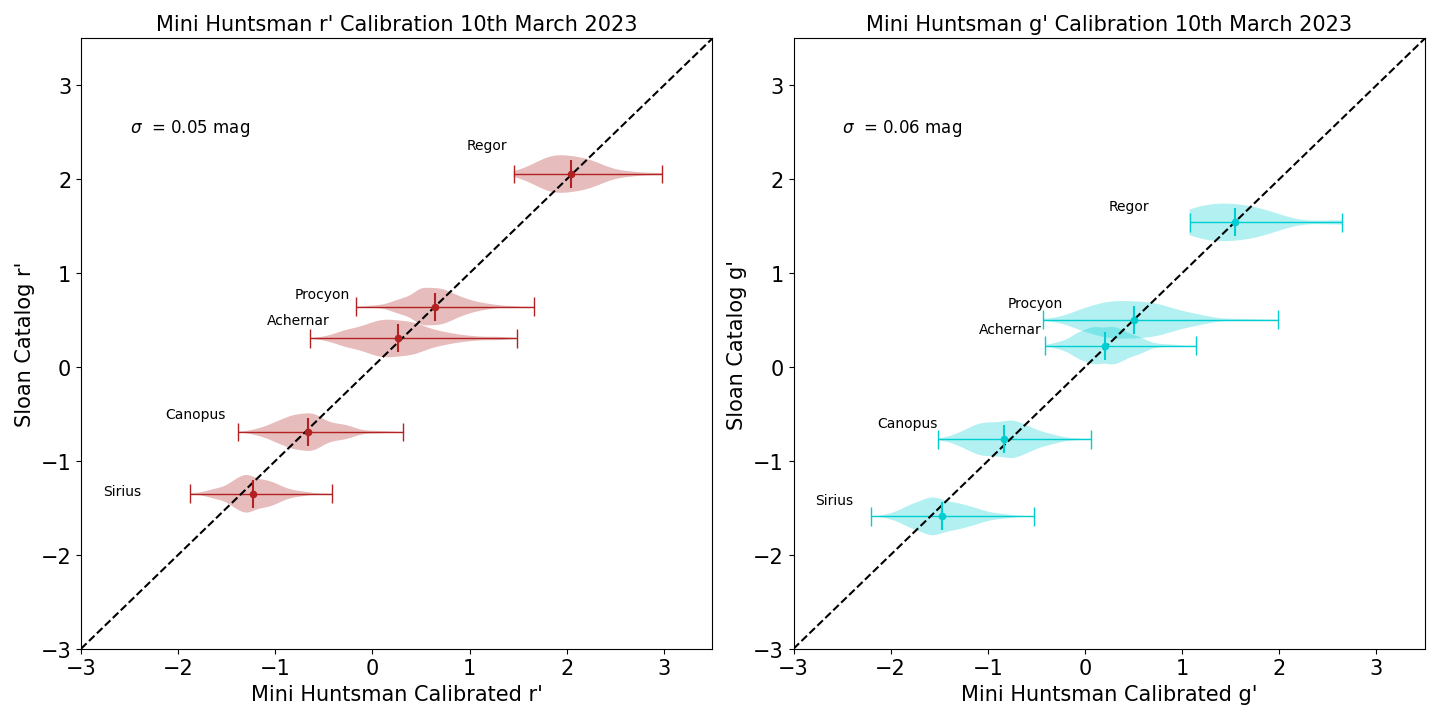}
    \caption{An example calibration plot for the 10th of March, showing the calibrate mini-Huntsman magnitude as a function of calculate Sloan catalog magnitude for 5 bright reference stars. Observing logs show a very windy day, passing cloud, with tops of $28^{\circ}C$ ($82.4^{\circ}F$). A violin plot is chosen to illustrate the distribution of calibrated magnitudes over multiple sets of 1000 exposures for each target. The photometric error for this day is found to be 0.05 for $r\text{\text{\textquotesingle}}$ and 0.06 for $g\text{\text{\textquotesingle}}$, around the order of magnitude of the flat field error.}
    \label{fig:calib_plots}
\end{figure*}
\\
To account for differences between the mini-Huntsman Sloan filters, and the calculated Sloan catalog magnitudes, the colour indices are calculated using data from the 8th of March 2023 which contains the largest sample of stars of varying colours of 24 target stars, varying from a B-V colour of -0.23 to 1.84. Following the results of \cite{nickel_daylight_2021} we re-calculate the zeropoint, extinction coefficient for every day of observations. Sydney is a large city, with approximately 5.3 million people located on the coast in a geographical basin. Aerosols from city pollution and changes in humidity and air density are frequent and persistent, which motivates us to perform daily calculations of the extinction coefficient following the results of \cite{nickel_daylight_2021}. We find that the median extinction coefficient for $k_{r\text{\text{\textquotesingle}}} = 0.22 \pm 0.12$ and $k_{g\text{\text{\textquotesingle}}}=0.30\pm0.19$. This is similar to the results of \cite{nickel_daylight_2021} who finds extinction values for V band observations of reference stars to be $k_{V} = 0.25 \pm 0.08$ to $k_{V} = 0.30 \pm 0.07$ for the months of February to April and May to July respectively. The higher standard deviation in our calculated extinction coefficient, may reflect more variable atmospheric conditions than observed by \cite{nickel_daylight_2021} in central Europe. This is consistent with our observing site in Sydney being located close to the coast, at a lower altitude of 61m as opposed to 200m. Similarly to \cite{nickel_daylight_2021}'s results, we don't find a correlation between photometric error and the calculated extinction coefficient per day.  
\\
\\
\autoref{fig:airmass_plots} illustrates an example of a typical day of observations on the 31st of March 2023, including 5 bright reference stars for photometric calibration of Betelgeuse observations. Each data point is calculated as the median of a set of 1000 exposures, with errorbars calculated as the estimated flat field error of 0.05 mag, and statistical error in quadrature. The Chi squared value shows a good fit, with 0.93 for $r\text{\text{\textquotesingle}}$ and 0.88 for $g\text{\text{\textquotesingle}}$. We note that scatter about an airmass of 1 corresponds to the same time at which lawn mowing was reported close to the observatory, picking up dust in the sky which could be a source of deviations from the linear trend that cannot be explained by the error bars in $g\text{\text{\textquotesingle}}$. The extinction coefficient for this day is found to be 0.27 for $r\text{\text{\textquotesingle}}$ and 0.54 for $g\text{\text{\textquotesingle}}$ with zeropoints of 21.65 and 22.73 respectively. Scatter about an airmass of 1 particularly in $g\text{\text{\textquotesingle}}$ may also explain a higher extinction coefficient than is calculated for the median over the several month period.
\\
\\
To quantify uncertainties in due to zeropoint calibration, we calculate the standard deviation of the difference in flux between the calibrated mini-Huntsman magnitudes and the catalog Sloan magnitudes for each day of observations. We also observe a similar magnitude limit for photometric calibration stars as \cite{nickel_daylight_2021} of V band 3 mag. \autoref{fig:calib_plots} illustrates the final fit for 5 bright reference stars on an typical example day, the 10th of March 2023. Calibrated mini-Huntsman magnitudes are shown to be in agreement with catalog Sloan magnitudes, with a standard deviation of 0.05 mag in $r\text{\text{\textquotesingle}}$ and 0.06 mag in $g\text{\text{\textquotesingle}}$. Violin plots show the distribution of data across multiple sets of 1000 exposures throughout the day, and the central data point is the median of this distribution. 
\\
\\
Across all days, the error is found to be a median of $0.05 \pm 0.03$ in $r\text{\text{\textquotesingle}}$ and $0.07 \pm 0.06$ in $g\text{\textquotesingle}$. These errors are comparable to the $0.05$ maximum predicted error due to the flat field uncertainty, and are also comparable but higher than the V band error found by \cite{nickel_daylight_2021} of $0.02 \pm 0.01$ mag (February-April) and $0.04 \pm 0.01$ mag (May-July). This higher error, and high standard deviation in the error across multiple days likely reflects our flat field error, the sporadic number of observed stars, as well as varied star colours and magnitudes in our preliminary observations as opposed to the consistent survey conducted by \cite{nickel_daylight_2021}. We also note that \cite{nickel_daylight_2021} prioritises the photometric accuracy of Betelgeuse observations by observing at optimal daytime conditions (lowest Sun altitudes) and during afternoon/twilight to prioritise photometric accuracy of Betelgeuse observations. In comparison, we have deliberately sampled a wide range of daytime observing conditions to understand the typical upper limit on photometric errors from data utilising the entire day. 
\\
\\
Errors due to transformation between photometric systems, zeropoint errors arising from number and variety of target stars as well as sampling different airmasses, and calibration uncertainties like the flat field error have been identified so far. Many of these factors may be improved upon in future work to mitigate the impact to the total photometric accuracy. Parameters that impact the atmospheric scintillation at high altitudes, local seeing conditions, Poisson noise, and factors influencing the overall SNR of observations are now considered in the context of daytime observations. 

\subsection{Poisson and Scintillation Noise}\label{noise}
Atmospheric scintillation noise originating in the upper atmosphere and Poisson noise from discrete photon measurements are two important sources of noise that contribute to the overall photometric error of daytime observations. Following the method of \citep{nickel_daylight_2021}, we estimate the median Poisson noise SNR ($SNR_{p}$) for each set of observations during the same epoch where the sky background is much brighter than the read noise and dark current:
\begin{equation*}
SNR_{p} = \frac{f_{r*}\times t}{[(f_{r*}\times t)+(\rho_{sky}\times t \times n_{pix})]^{1/2}}
\end{equation*}
Where $f_{r*}$ is the source rate, $\rho_{sky}$ is the sky rate, $n_{pix}$ is the number of pixels in the source aperture, and $t$ is the exposure time. The scintillation SNR ($SNR_{sc}$) for a set of observations during the same epoch can be expressed as:
\begin{equation*}
SNR_{sc} = \frac{MED(f_{r*}\times t)}{{\sigma (f_{r*}\times t)}}
\end{equation*}
Where $MED(f_{r*})$ is the median of the total source counts for the target in an aperture in units of $e^{-}$ and $\sigma (f_{r*})$ is the standard deviation of total source counts if there was no other source of noise present \citep{nickel_daylight_2021}. Because the contribution of scintillation noise and Poisson noise cannot be separated from observations, we use the normalised scintillation index $\sigma_{sc}$ defined as:
\begin{equation*}
\sigma_{sc} = \frac{1}{SNR_{sc}}
\end{equation*}
We take the total SNR of the observations to be the median $SNR_{p}$ of a set of observations. We also define another parameter, the detection probability, as the percentage of attempted detections to successfully detected sources. This parameter will be a function of the algorithm and source detection parameters used, as well as the environmental conditions on the day of observation.
\\
\begin{figure}[ht!!!]
    \centering
    \includegraphics[width=\linewidth]{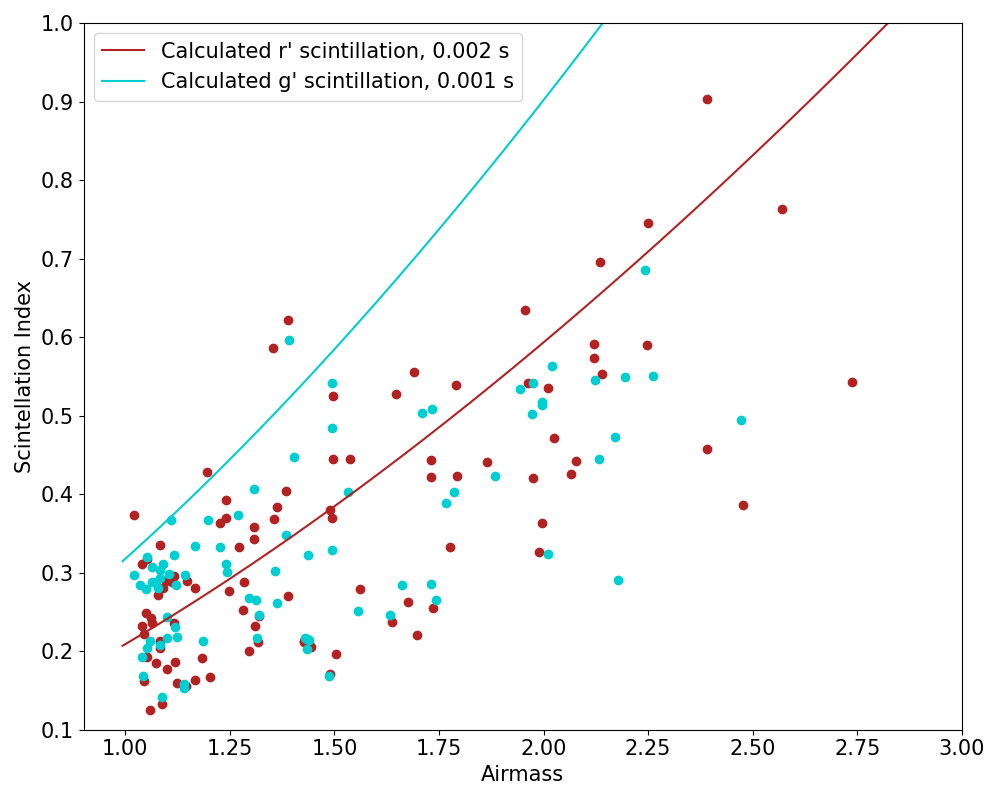}
    \caption{The scintillation index for $g\text{\textquotesingle}$ and $r\text{\textquotesingle}$ observations as a function of airmass, plotted together with the calculated theoretical scintillation index using Young's approximation for the median exposure time for each filter. We find that the $r\text{\textquotesingle}$ observations are in general agreement with Young's approximation, however Young appears to overestimate the scintillation for shorter $g\text{\textquotesingle}$ exposures. However, this could also be due to a bias where by faster exposures times with lower SNR and higher sky Poisson noise in $g\text{\textquotesingle}$ are less likely to be detected reliably.}
    \label{fig:scint}
\end{figure}
\\
By inspection of calculated Pearson's correlation coefficients, we find the scintillation index for exposures taken at the optimal exposure time for each target and sky brightness is strongly correlated solely with the airmass of observations with a correlation coefficient of 0.64. In comparison, examined environmental parameters of Sun altitude, Sun separation, and sky background intensity have correlation coefficients of -0.07, -0.06 and 0.19 respectively. The theoretical scintillation is calculated using a basic Young's approximation given by:
\begin{align*}
    \sigma^{2} = 10 \times 10^{-6} D^{-3/4} t^{-1}(cos\gamma)^{-3} exp(-2h_{obs}/H)
\end{align*}
Where D is the diameter of the telescope in meters, t is the exposure time, $h_{obs}$ is the height of the observatory in meters (64m for Macquarie Observatory), $\gamma$ is the zenith distance of the target observation, and H is the scale height of the atmosphere, generally considered to by 8000m \citep{osborn_atmospheric_2015}. In \autoref{fig:scint} we plot the measured scintillation index as a function of airmass for $g\text{\textquotesingle}$ and $r\text{\textquotesingle}$, as well as the calculated scintillation index for the median exposure time used for $g\text{\textquotesingle}$ and $r\text{\textquotesingle}$ of 0.001s and 0.002s respectively. We find excellent agreement with Young's approximation at these wavelength and exposure times for the $r\text{\textquotesingle}$ exposures, however we find that this is overestimated for $g\text{\textquotesingle}$ exposures. This may be due to the fact that exposures in $g\text{\textquotesingle}$ have a shorter exposure time as well as a higher sky surface brightness and thus lower SNR leading to a poor detection probability. As a result, collected data may be biased toward lower scintillation indexes at lower airmass. 
\\
\begin{figure*}[ht!!!]
    \centering
    \includegraphics[width=11cm]{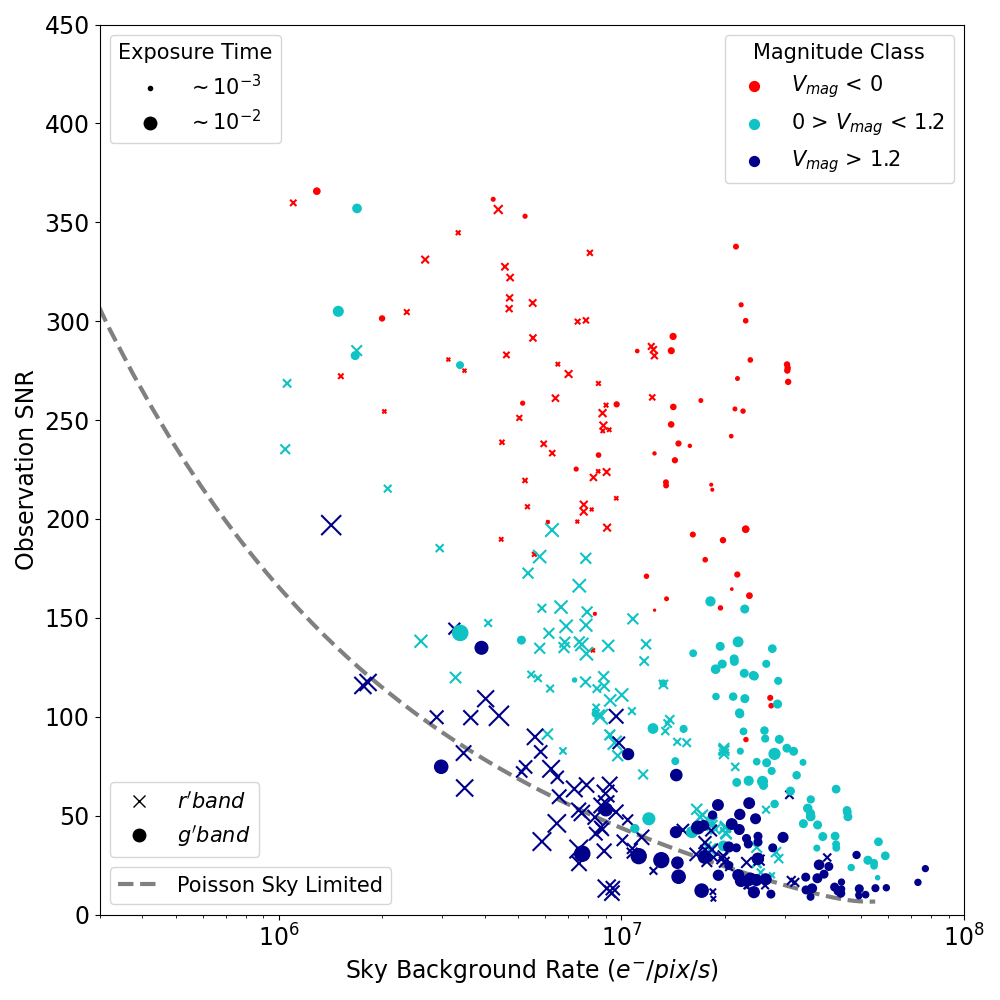}
    \caption{The the SNR of the target observation as a function of the sky background rate in $e^{-}/pix/s$. Observations are split up into magnitudes classes of roughly equal sample sizes of $V_{mag} \leq 0$, $0 > V_{mag} < 1.2$ and $V_{mag} \geq 1.2$. The size of the markers indicate the exposure time of the observation. The shape of the marker indicates the bandpass. For the faintest class of stars, the exposure time decreases as a function of the sky background rate. The calculated trend for Poisson sky noise limited observations is plotted, and is found to describe the data in the faintest magnitude class, indicating that observations of stars with $V_{mag} < 1.2$ are Poisson sky noise limited during the day. In contrast, stars in the brightest magnitude class are found to be scintillation noise dominated.}
    \label{fig:SNR_sky}
\end{figure*}
\\
\noindent
We now explore the Poisson noise from the sky and target $SNR_{p}$ as a function of the measured sky surface brightness. Observed targets are separated into roughly equal magnitude classes each with $\sim 120$ sets of observations to examine the noise properties of targets of different magnitudes. These classes are $V_{mag} < 0$, $0 > V_{mag} < 1.2$ and $V_{mag} > 1.2$. \autoref{fig:SNR_sky} shows the sky background rate in $e/pix/s$ as a function of SNR for these observations. For the faintest stars, the maximum exposure time is limited by the sky background, rather than the target flux. Fitting a linear relationship to \autoref{fig:exp_times}, we fit a line corresponding to the SNR of a Poisson sky noise limited case of the form:
\begin{equation*}
    SNR = \frac{f_{r*}\sqrt{t}}{\sqrt{f_{sky}}}
\end{equation*}
Where $f_{r*}$ is the target rate, $f_{sky}$ is the sky background rate, and $t$ is the exposure time. For the fainted magnitude class, this relationship is able to describe the data, and so we can conclude that observations of stars fainter than $\sim 1.2 V_{mag}$ are likely to be Poisson noise dominated during the day. For stars brighter than $\sim 0$ $V_{mag}$, we find that they do not follow a Poisson noise dominated trend. By examining the Pearson's correlation matrix with our observed parameters, we find that the SNR of these observations most closely correlated with scintillation index, and approximates a linear relationship.   
\begin{figure*}[ht!!!]
    \centering
    \includegraphics[width=\linewidth]{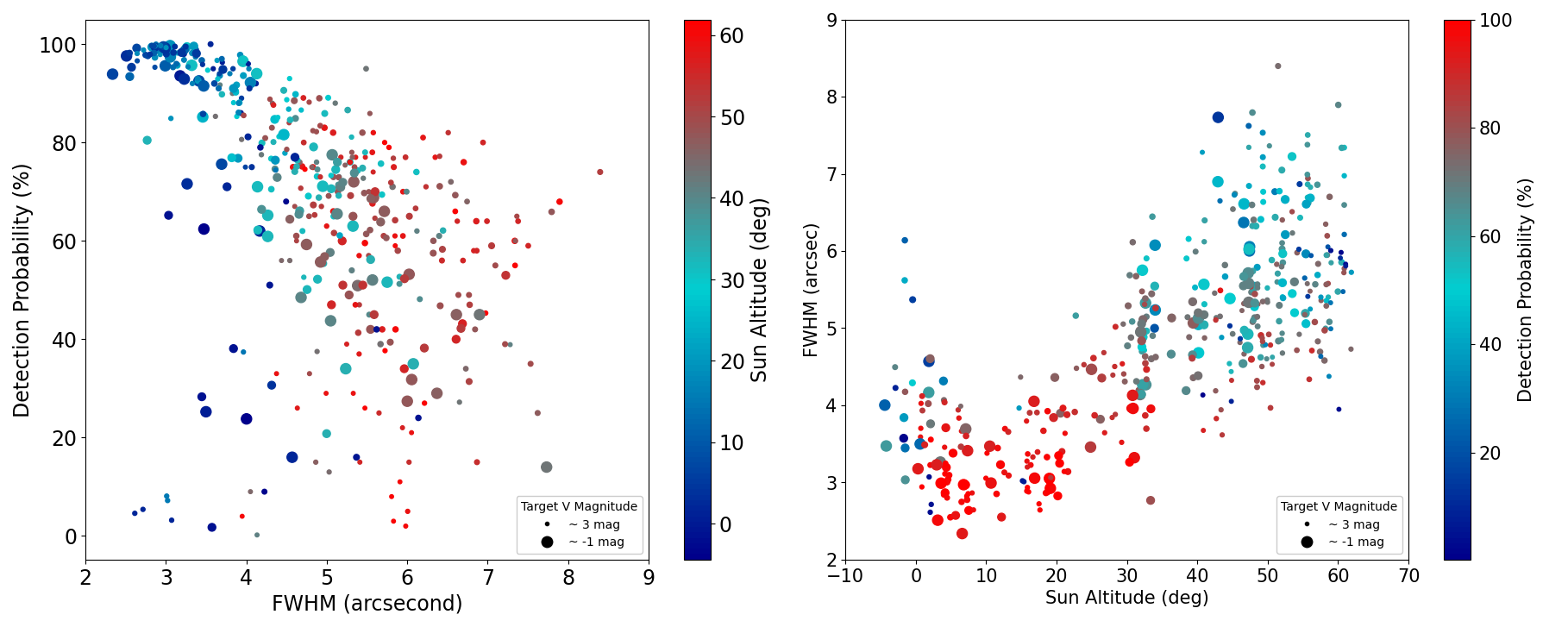}
    \caption{Left panel: The detection probability as a function of the measured FWHM, with colour bar as the Sun altitude in degrees at the time of observation, and the target magnitude is the relative size of the data points. Detection probability is show to decrease with increased FWHM corresponding to the local seeing conditions, and Sun altitude. Right panel: We plot the FWHM as a function of the Sun altitude with detection probability as the colour bar. the  We see a relationship between the FWHM and the detection probability, with observations at a detection rate greater than 80\% occurring in majority below a FWHM of 4 arcseconds, and a Sun altitude of 30 degrees.}
    \label{fig:detect_probs}
\end{figure*}
\subsection{Detection Probability and Local Seeing Conditions}

We now consider the detection probability per set of exposures, and the local seeing conditions. The local seeing can be described by the measured FWHM of the observation, which we find to be related to the standard deviation of the movement of the target centroid. \autoref{fig:detect_probs} illustrates the relationship between the Sun altitude at the time of the observation, the FWHM and the detection probability. In the left most panel, we find that the detection probability generally decreases with an increase in FWHM indicating poorer local seeing conditions, and increasing Sun altitude is generally correlated with an increase in the FWHM, and thus a degradation of the seeing conditions. We do however see a few outliers which have a small FWHM and lower detection probability which are explored in the right most panel. Here we see that after sunset, the seeing conditions worsen again, with an increased FWHM and decreased detection probability. This is consistent with the fact that the local seeing conditions are impacted by the temperature equilibrium of the observing site \citep{sterken_atmospheric_1992}. Large temperature gradients resulting in turbulent cells and convection in the air immediately above the observing site may be the cause of the degradation of seeing conditions on sunset and approaching midday, however more data into the evening, and site temperature information is needed to confirm this. From these results, we see that the best time to observe with the best seeing conditions and the highest detection probability is between sunrise, and $\sim 30$ degrees Sun altitude resulting in a $> 80\%$ detection probability. We also find that the seeing conditions are wavelength dependant, with observations in $g\text{\textquotesingle}$ found to have a $0.4 \pm 0.2$ arcsecond larger PSF FWHM than those in $r\text{\textquotesingle}$ at fixed Sun Altitude. This is expected, as the effects of seeing on the FWHM is proportional to the wavelength of light by $\lambda ^{-1/5}$.  
      
\subsubsection{Daytime Sky Surface Brightness and Colour}\label{sky}

\begin{figure*}[ht!!!]
    \centering
    \includegraphics[width=\linewidth]{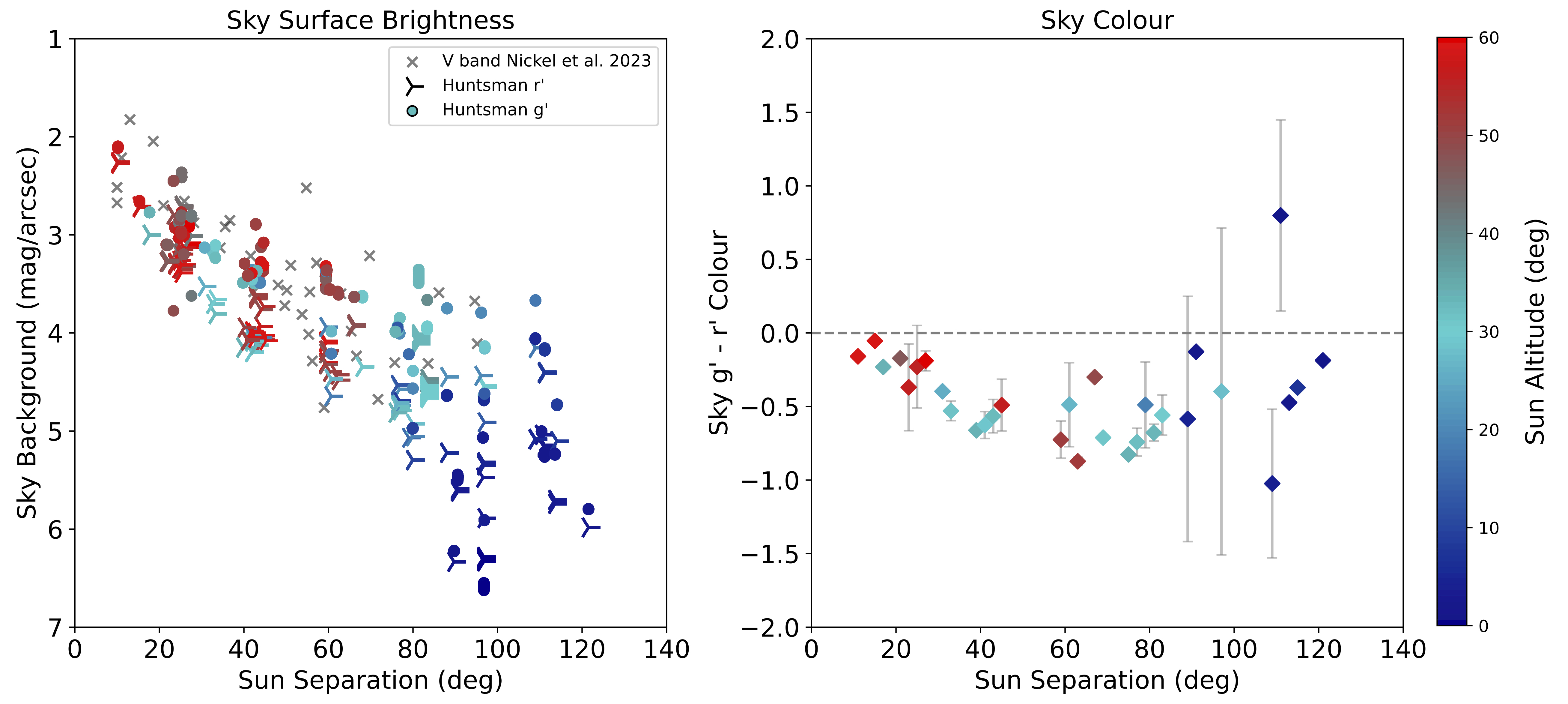}
    \caption{Measured sky surface brightness is plotted as a function of Sun separation from the observed target for $g\text{\textquotesingle}$ and $r\text{\textquotesingle}$. Grey crosses are $V$-band filter observations from \cite{nickel_daylight_2021}. Our observations are in general agreement with the literature, approaching 2 mag/arcsec for Sun separation $> 30$ degrees. Sky background surface brightness gradually decreases until falling sharply for Sun altitudes $< 10$ degrees, and Sun separation $> 100$ degrees. The sky is brighter in $g\text{\textquotesingle}$ than $r\text{\textquotesingle}$ at Sun separation $< 90$ degrees and $> 40$ degrees. $g\text{\textquotesingle}$ - $r\text{\textquotesingle}$ sky colour is explored as a function of Sun separation with the colour bar representing Sun altitude. Data is binned via Sun separation. At Sun separations of $\sim 20$ degrees or higher the sky is redder in colour, and gradually becomes more bluer at larger Sun separations. Sky colours cover a broader range at lower Sun altitudes and larger Sun separations.}
    \label{fig:sky_bright}
\end{figure*}

The brightness and colour of the daytime sky is an important factor influencing the utility of optical daytime astronomy. In \autoref{fig:sky_bright} we present the measured sky background in mag/arcsec for $g\text{\textquotesingle}$ and $r\text{\textquotesingle}$ as a function of Sun separation and Sun altitude, alongside observations in V from \cite{nickel_daylight_2021}. 
\\
\\
We find agreement with the results of \cite{nickel_daylight_2021}. The sky brightness approaches 2 mag/arcsec in $g\text{\textquotesingle}$ and $r\text{\textquotesingle}$ for Sun separation $< 30$ degrees and high Sun altitudes. Sky brightness drops to a minimum at the largest Sun separation distances, and smallest Sun altitudes. A break in the relationship occurs for Sun altitudes approaching 90 degrees, where the sky brightness drops to 7 mag/arcsec in $g\text{\textquotesingle}$ and $r\text{\textquotesingle}$ near Sunset. We present the sky colour as a function of Sun separation and Sun altitude in \autoref{fig:sky_bright}. In this plot, we bin the data by Sun separation to determine the $g\text{\textquotesingle} - r\text{\textquotesingle}$ sky colour, and error bars are used to represent the spread of the data in the bins as the standard deviation. For smaller Sun separation angles, and higher Sun altitudes, we find the sky colour approaches zero, or whiter in colour. This is expected, as path length through the atmosphere is shorter and so less blue light can be scattered. At larger Sun separation angles the sky becomes bluer again due to the longer path length and increased Rayleigh scattering at these angles. Approaching lower Sun altitudes and the largest Sun separation angles, we see a splitting in the colour and increased scatter. This is due to the sunset where the sky becomes redder in colour in some directions, and bluer in others depending on the time and azimuth of the observation.

\subsection{Observations of Betelgeuse}

\begin{figure*}[ht!!!]
    \centering
    \includegraphics[width=14cm]{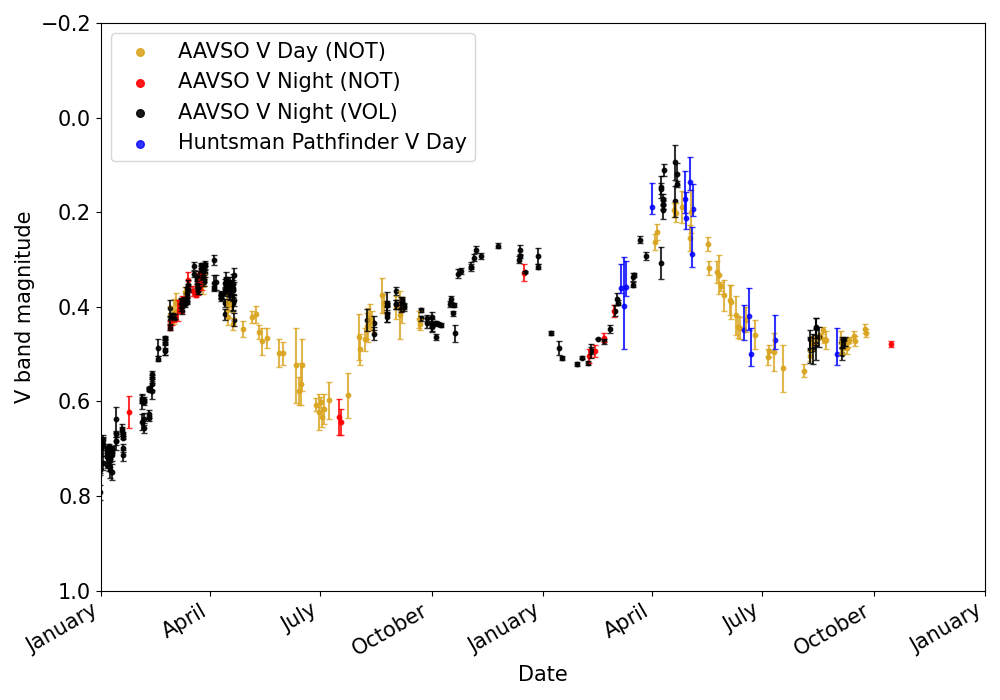}
    \caption{Huntsman Pathfinder calculated V band observations plotted in blue, with current \cite{nickel_daylight_2021} (observer code NOT) observations during the day in yellow, and during the night in red. In addition, AAVSO (observer code VOL) observations of Betelgeuse are also included for reference in black taken at night. Asymmetrical error bars are calculated to accommodate the flat field error which tends to dominate. A period of consistently poor weather between May and June in the transition from Autumn to Winter resulted in a large gap in our data. Despite this, our results show general a good agreement with both \cite{nickel_daylight_2021} and AAVSO observations, with comparable errors.}
    \label{fig:bj_obs}
\end{figure*}

We demonstrate the capability of the Huntsman Telescope operating as a optical daytime facility by monitoring the bright ecliptic variable star Betelgeuse. Here we present our light curve over the course of a 7 month period from March through to September. The median observed magnitude of Betelgeuse is calculated for each day from thousands of observations over multiple epochs in $g\text{\textquotesingle}$ and $r\text{\textquotesingle}$, and the V band magnitude is calculated to compare with AAVSO observations using the equations of \cite{jester_sloan_2005} as follows:
\begin{equation*}
    V = g\text{\textquotesingle} - 0.59\times(g\text{\textquotesingle}-r\text{\textquotesingle}) - 0.01 \pm 0.01
\end{equation*}
Similarly to the reference stars, we employ the same method of data reduction as described in \autoref{sec:method}. In order to compute the error in our Betelgeuse ($\sigma _{BJ}$) measurement from differential photometry, we estimate the error including the reference stars ($\sigma _{Ref}$) as follows:
\begin{equation*}
    \sigma _{Lower} = \sqrt{\sigma _{BJ} ^{2} + \sigma _{Ref} ^{2}}
\end{equation*}
\begin{equation*}
    \sigma _{Upper} = \sqrt{\sigma _{BJ} ^{2} + \sigma _{Ref} ^{2} + \sigma _{Flat} ^{2}}
\end{equation*}
We calculate asymmetrical errorbars, because the flat field error that dominates ($\sigma _{Flat}$) is considered to be a lower limit on the magnitude of the target star as the transmission of the vignette pattern can only decrease the flux from the target. \autoref{fig:bj_obs} presents our Betelgeuse light curve from our preliminary results. mini-Huntsman data in blue shows general good agreement with AAVSO data taken by \cite{nickel_daylight_2021} during the day and night (yellow and red points respectively), as well as AAVSO data at night (black points). Unfortunately a period of consistently poor weather between May and June in the transition from Autumn to Winter resulted in a large gap in our data. Despite this, our measurements show agreement with existing measurements, with comparable error bars. Again we note that the observations of \cite{nickel_daylight_2021} during the day where the taken at the most opportune conditions as the object of these results was a reliable Betelgeuse light curve. Our data is taken in a variety of poor and favourable daytime only conditions to explore the parameter space, and thus the quality of our data is likely to reflect this.
\subsection{International Space Station Observations}
As part of our initial investigations into performing photometry of bright satellites for the purposes of SSA with the Huntsman Telescope, we also present a preliminary analysis of an International Space Station (ISS) pass in $g\text{\textquotesingle}$. The images are taken using the full frame to ensure the greatest probability of capturing the ISS in shot due to the know uncertainties in available TLE's, as well as to account for any misalignment in the pointing model. As a result, these images are able to be flat field corrected. We median combine 114 $g\text{\textquotesingle}$ flat field images taken during twilight to create a master flat and apply this to each science frame. Only photometric errors are considered and plotted for satellite targets, due to their dynamic movement through different sky conditions. To determine the rough target brightness, a fixed aperture size of 20 pixels is used, and reference stars are used to calibrate the magnitude of the source. \autoref{fig:iss_pass} shows the calculated magnitude of the target as a function of local time (AEST) on the left and the calculated satellite trajectory on the right. 
\\
\\
The target is tracked successfully using Software Bisque's TheSkyX from the observer horizon above the tree line, until approximately 46 degrees in altitude where the target is lost and drifts out of frame. We anticipate this is either due to TLE errors, pointing model errors, possible limitations of the mount tracking speed, or a combination of all three. The target is reacquired again as it begins to descend until the mount reaches the meridian where tracking is stopped due to the mechanical limitations of the equatorial mount. The target is well resolved, with major features such as the solar panels, radiators, and station modules easily visible at the highest altitude in the pass. It is measured at a maximum brightness of $g\text{\textquotesingle}$$\sim$ 1.2 mag.

\section{Discussion}\label{sec:discusssion}
We now compare our results with that of the literature, and discuss the impacts of our derived detection limits, photometric accuracy and observing conditions on the productivity of an optical daytime observing facility. 

\subsection{Photometry and Detection Limits}
\begin{figure*}[ht!!!]
    \centering
    \includegraphics[width=\linewidth]{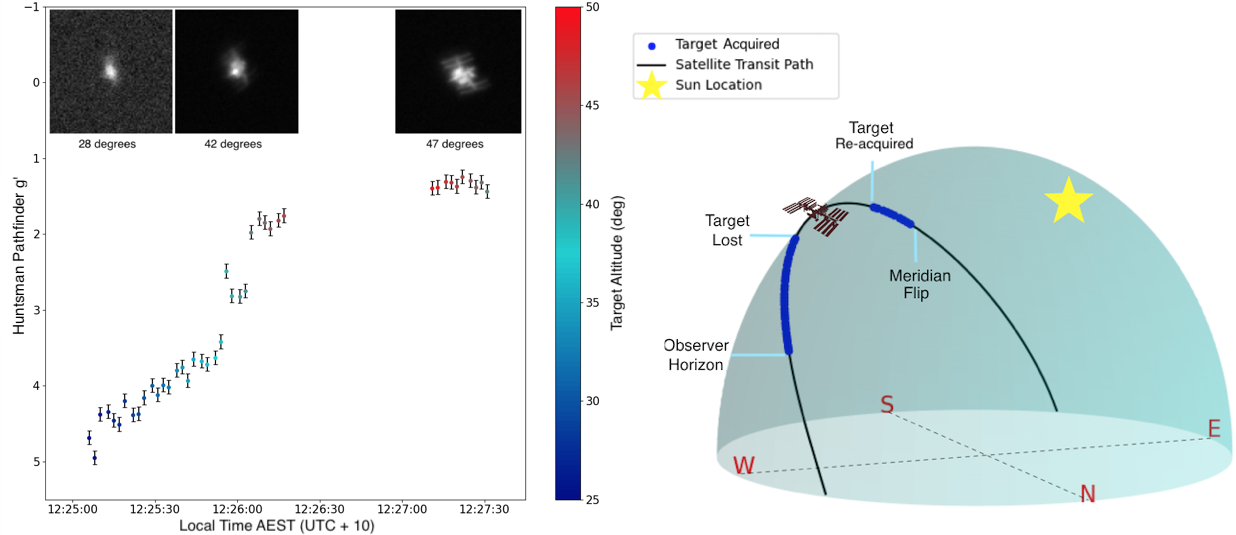}
    \caption{The trajectory (right panel) and brightness (left panel) of the ISS in $g\text{\textquotesingle}$ taken by the Huntsman Pathfinder. The error bars are calculated as the photometric error from the calibration stars used on the day, and are flat field corrected. The colour is the target altitude. The target increases in brightness until reaching a maximum of $g\text{\textquotesingle}$ 1.2mag. It is lost for a short period during the transit when it drifts out of the FOV, either due to errors in the TLE, the pointing model, or limitations of the mount tracking speed. It is reacquired as it begins the decent, before the mount stops tracking at the meridian due to the mechanical limitations of the equatorial mount.}
    \label{fig:iss_pass}
\end{figure*}

Across all observations over 7 months we find a median photometric error of $0.05 \pm 0.03$ in $r\text{\textquotesingle}$ and $0.07 \pm 0.06$ in $g\text{\textquotesingle}$. Our errors are higher than previous results by \cite{miles_daytime_2007, nickel_daylight_2021} who report $ \pm 0.03$ V band and $0.02 \pm 0.01$ mag to $0.04 \pm 0.01$ mag respectively. This likely reflects the variation in observing conditions, and number of reference stars day to day in this initial exploratory survey. It is also comparable to our considerable calculated flat field uncertainty of $\pm 0.05$ mag which will be corrected in future works. We also find a 0.03 mag increase in the standard deviation of the photometric error for $g\text{\textquotesingle}$ compared to $r\text{\textquotesingle}$, which may be as a result of poorer seeing conditions at shorter wavelengths with $g\text{\textquotesingle}$ found to have a median $0.4 \pm 0.2$ arcsecond larger PSF FWHM than $r\text{\textquotesingle}$ observations. A bluer sky colour approaching a $g\text{\textquotesingle}$ - $r\text{\textquotesingle}$ colour of -1 at sun separation of 60-80 degrees as well as a $\sim 1$mag increase in sky brightness in $g\text{\textquotesingle}$ may also be a cause of lower target SNR in $g\text{\textquotesingle}$. 
\\
\\
As shown in \autoref{fig:bj_obs}, our light curve of Betelgeuse is in agreement with those found on the AAVSO archive both for daytime and night time observations. This agreement demonstrates the Huntsman Telescope's potential as an op observatory dedicated to observing bright, variable ecliptic stars like Betelgeuse. 
\\
\\
Our preliminary results of tracking and imaging the ISS demonstrates the potential for using the Huntsman Telescope facility for the purposes of SSA during the day. In \autoref{fig:iss_pass} the target is acquired at an altitude of $\sim 25$ degrees and tracked through to the meridian, reaching a maximum brightness of $g\text{\textquotesingle}$$\sim$ 1.2 mag. These results demonstrate that the ISS is clearly visible during the day, and photometry of SSA targets is possible. 
\\
\\
The photometric accuracy achieved by mini-Huntsman are smaller than that which is required to differentiate between satellites of different composition. For example, the study conducted by \cite{schmitt_multicolor_2016} show that errors of 0.2 mag can yield useful insight. Our errors of $0.05 \pm 0.03$ in $r\text{\textquotesingle}$, are within this threshold. However $0.07 \pm 0.06$ in $g\text{\textquotesingle}$ will require further exploration to reduce the standard deviation of photometric accuracy across all observing conditions explored in this work. If similar calibration campaigns were employed for SSA work it may unlock a new regime of classifying bright satellite passes during the day using their relative brightness at different wavelengths of light. 
\\
\\
We report an average 10-sigma detection limit across all exposures and environmental conditions of $\sim 4.9$ mag V band, with the faintest star reliably detected using our detection algorithm being HIP99120 with a V band magnitude of 4.94 at a sun altitude of 10 degrees. For midday observations we report a detection limit of $\sim 4.6$ mag V observing HIP3245 at a Sun altitude of 62 degrees. We report visually confirming observations of stars up to V band 6 mag at midday, however they were not reliably detected using our source detection algorithm.
\\
\\
A detection limit of 4.6 mag during the day means the number of bright stars observable with mini-Huntsman is less than 900, estimated using the Tycho catalog, or $\sim 0.001\%$ of the stars visible to the Huntsman Telescope at night using a comparable filter, single lens, and similar exposure time. It is clear that this magnitude limit restricts the practicality of observing variables stars during the day, however this observing mode may still be useful for select cases. Precise and long term photometry of stars such as Aldebaran and Pollux may help to determine if these stars have planets or long secondary period variations \citep{reichert_precise_2019}. In addition to Betelgeuse, long period variable legacy stars monitored by AAVSO \citep{hutton_aavso_2009} such as 
eclipsing binary VV Cephei (V 4.90 mag), recurrent nova system RS Ophiuchi (V 5.5 mag peak brightness, AAVSO), binary star Sigma Geminorum (V 4.29 mag) and multiple star system Beta Persei (V 2.12) among others may be potential candidates for the Huntsman Telescope system during the day. Continuous, long term monitoring of these targets in addition to other program star targets monitored AAVSO may be potential targets for daytime observing programs. 
\\
\\
A detection limit of $\sim4.6$ mag at midday allows for a large sample of photometric calibration stars in order to perform SSA photometry during the day. However the number of observable satellites may potentially be limited to only 4, according to NORAD catalogs \citep{kaminski_optimizing_2021}. However, there is some uncertainty around the brightness of potential SSA targets during the day, with Starlink satellites of particular interest \citep{cole_sky_2020, kaminski_optimizing_2021, zimmer_overcoming_2021,fankhauser_satellite_2023}. These satellites are reported to reach magnitudes brighter than 5th to 6th magnitude during terminator conditions. In addition there is some uncertainty on their Earthshine illuminated brightness, with one report from \cite{zimmer_overcoming_2021} detecting them up to 2.6 mag during the day. If this were verified, the number of satellites with the current setup could potentially increase to thousands. Future works with the Huntsman Telescope system will focus on attempting to detect these Starlink satellites during the day and compare them to terminator illumination conditions as well as evening conditions such as those explored in  \cite{steindorfer_daylight_2020}. 

\subsection{Daytime Observing Conditions}
Over the course of 7 months we have explored the observing conditions for our site at Macquarie Observatory during the day, and report the impact of environmental conditions on the quality of observations. We find that an increase in scintillation noise is only correlated with airmass, and is not dependant upon conditions such as Sun altitude, Sun separation, and sky surface brightness. In \autoref{fig:scint} we find that Young's approximation describes the trend with airmass in $r\text{\textquotesingle}$, but overestimates the observed scintillation in $g\text{\textquotesingle}$. This may be due to bias in detection portability for higher scintillation noise dominated observations, however more data is needed to explore this hypothesis. We find that by splitting the data up into rough magnitude classes of equal sample size $V_{mag} \leq 0$, $0 > V_{mag} < 1.2$ and $V_{mag} \geq 1.2$ that the SNR of observations of targets with $V_{mag} \leq 0$ are linearly with the scintillation index. From these results we conclude that only observations of the brightest stars are scintillation noise dominated. In contrast, we find that observations of stars $V_{mag} \geq 1.2$ follow a theoretical Poisson noise dominated curve. These findings illustrate the fundamental limit to optical observations during the day. While observations of 0th magnitude stars and brighter may be achieved with multiple observations to sample the scintillation noise, observations of stars fainter than $\sim$ 1st magnitude are Poisson sky noise dominated with our telescope. Overcoming Poisson sky noise may be achieved by reducing the contribution of the sky through the use of filters and polarises, or by changing the optical set up by decreasing the pixel scale or increasing the telescope aperture. 
\\
\\
As explored in \autoref{fig:SNR_sky} we find that the detection probability is impacted by the local seeing conditions. The majority of detection probabilities of greater than 80\% occur between Sun altitude of 0 and 30 degrees, with FWHM lower than 4 arcsec. This may indicate that Winter observing conditions in Australia could be more favourable for daytime observations where Sun altitudes are typically around $\sim 33$ degrees (Winter solstice). FWHM in $g\text{\textquotesingle}$ compared to $r\text{\textquotesingle}$ is larger, with median PSF FWHM of $4.2 \pm 2.0$ arcsec and $3.7 \pm 1.1$ arcsec respectively. Considering only calibration stars with a verity of colours and magnitudes, the detection probability is roughly equal in $g\text{\textquotesingle}$ and $r\text{\textquotesingle}$ of $72\% \pm 28\%$ and $71\% \pm 24\%$ respectively. Seeing is most poor immediately before or after Sunset, consistent with periods of greatest thermal fluctuation. Seeing also deteriorates during the middle of day, at the highest Sun altitudes. Investigations into whether this impact can be reduced by comparing conditions at the Macquarie Observatory location to Siding Spring Observatory, as well as the impact of direct Sunlight as opposed to housing the instrument in a dome for daytime observing will be explored in future works.    
\section{Conclusions and Future Work}
In conclusion we find that a photometric accuracy of 1\% to 10\% is possible with the Huntsman Telescope system. Local seeing conditions are found to be impacted by Sun altitude, with the best observing conditions taking place between Sun rise, to a Sun altitude of 30 degrees, with the majority of observations of all stars achieving a detection probability of 80\% or better, and FWHM below 4 arcsec. 
\\
\\
Overall we find that the FWHM is smaller for observations at longer wavelengths, and on average produce higher SNR observations of target stars. Our detection limit using our current source detection algorithm for midday observations is found to be V band 4.6 mag, increasing to $\sim 4.9$ mag for observations in the afternoon under more favourable sky conditions. 
\\
\\
Scintillation is found to be well described during the day by a Young's approximation, particularly for longer explore times of 0.002 s. However we see better than expected scintillation at shorter exposure times, possibly due to bias in source detection probability at these exposure times. We find that observations of targets V band 0th mag and brighter are likely scintillation noise dominated, while stars fainter than V band 1.2 mag are Poisson sky noise dominated.  
\\
\\
Over the course of 7 months of daytime observations of Betelgeuse we find good agreement between our measurements and those recorded on the AAVSO archive by fellow Betelgeuse daylight observer \cite{nickel_daylight_2021}. Our photometric errors are dominated by our flat field error as the result of proprietary software bugs, which will be easily corrected in future observations.  
\\
\\
Our initial results exploring the tracking and photometry of satellites for SSA demonstrate the successful acquisition and tracking of the ISS. We present a light curve illustrating the change in brightness over the duration of the pass. We report difficulty tracking the station towards the highest point of the pass, possibly due to the tracking limits of the mount, or possible pointing errors. The photometric accuracy we can achieve for the ISS during the day potentially unlocks ways to characterise or assess the status of satellites passing overhead during the daytime.
\\
\\
As we move towards rolling out this new observing mode on the Huntsman Telescope, there are several improvements to be made in future works to optimise observing during the day. 
Firstly, for aperture photometry we will explore the use of a variable aperture size to account for a change in seeing conditions at different times during the day, and to maximise the SNR of observations. Autonomous focusing will be another point of interest in development, as variable temperatures during the day are found to be correlated with focusing offsets. In addition, we will also explore the use of long pass filters to maximise the SNR of both satellite and stellar targets. We may also be able to take advantage of the colour variation of the sky as a function of Sun separation to optimise filter selection for various observing conditions to maximise target SNR. Lucky imaging, whereby data is filtered to include only the most opportune moments with highest SNR \citep{smith_investigation_2009} is another technique that has not been explored in this work, and we intend to implement this in the next Huntsman Pathfinder daytime survey to maximise the quality of our observations. For the purposes of SSA we will continue to explore methods of increasing our detection limit via the use of targeted wavelength filters, as well as exploring the reported daytime magnitudes for Starlink satellites, in particular, taking into account Earthshine contributions. Finally, we will continue to monitor Betelgeuse during the day using the Huntsman Telescope facility, as well as exploring the feasibility of monitoring other long period variable stars of interest.

\begin{acknowledgement}
The authors would like to acknowledged the traditional owners of the land on which the Huntsman Telescope is situated, Gamilarray, Wiradjuri and Wayilwan Country. In addition we would like to acknowledged the traditional owners of the land on which the Macquarie University Observatory is located, the Wallumattagal Clan of the Dharug Nation – whose cultures and customs have nurtured, and continue to nurture, this land since time immemorial.
\\
\\
We would like to thank Canon Australia for their support of the Huntsman Telescope collaboration and Mini-Huntsman project. 
\\
\\
Finally the authors would like to acknowledge the generous assistance of Macquarie Observatory manager Adam Joyce who has provided continued technical support of this project that has made these results possible.\end{acknowledgement}

\paragraph{Software}

Astropy (The Astropy Collaboration 2013, 2018),
Scipy (Virtanen, P. et al, 2020),
PyMC3 (Salvatier, J. et al, 2016),
Photutils (Bradley, L. et al, 2016),
Source Extractor (Bertin, E. et al, 1996),
Skyfield (Rhodes, B. et al, 2019)

\paragraph{Author Contribution Statement}

\textbf{Sarah E. Caddy}: Conceptualization, Methodology, Software, Formal analysis, Investigation, Writing. \textbf{Lee R. Spitler}: Supervision, Conceptualization, Methodology, Review \& Editing. \textbf{Simon C. Ellis}: Supervision, Review \& Editing. 

\paragraph{Funding Statement}

S.C. and L.S. acknowledge support from an Australian Research Council Discovery Project grant \\
DP190102448

\paragraph{Competing Interests}

The authors report no conflict of interest to the best of our knowledge at the time of publication. 

\paragraph{Data Availability Statement}

Data collected with the Huntsman Pathfinder for the purposes of this work, and the python code used to reduce the data may be made available at the request of the first author. 


\printbibliography

\end{document}